# Quantum Robot: Structure, Algorithms and Applications

Dao-Yi Dong, Chun-Lin Chen, Chen-Bin Zhang, Zong-Hai Chen

Department of Automation, University of Science and Technology of China,

Hefei, Anhui, 230027, P. R. China

e-mail: dydong@mail.ustc.edu.cn

clchen@mail.ustc.edu.cn

**SUMMARY:** A kind of brand-new robot, quantum robot, is proposed through fusing quantum theory with robot technology. Quantum robot is essentially a complex quantum system and it is generally composed of three fundamental parts: MQCU (multi quantum computing units), quantum controller/actuator, and information acquisition units. Corresponding to the system structure, several learning control algorithms including quantum searching algorithm and quantum reinforcement learning are presented for quantum robot. The theoretic results show that quantum robot can reduce the complexity of O($N^2$) in traditional robot to O($N\sqrt{N}$) using quantum searching algorithm, and the simulation results demonstrate that quantum robot is also superior to traditional robot in efficient learning by novel quantum reinforcement learning algorithm. Considering the advantages of quantum robot, its some potential important applications are also analyzed and prospected.

## 1 Introduction

The naissance of robots and the establishment of robotics is one of the most important achievements in science and technology field in the 20[th] century [1-3]. With the advancement of technology [4-8], robots are serving the community in many aspects, such as industry production, military affairs, national defence, medicinal treatment and sanitation, navigation and spaceflight, public security, and so on. Moreover, some new-type robots such as nanorobot, biorobot and medicinal robot also come to our world through fusing nanotechnology, biology and medicinal engineering into robot technology. As viewed from the development tide of robots, intelligentization and micromation are two important directions. The key of intelligentization lies in improving the performance of sensors and increasing the speed of learning and behavior decision, and ceaseless micromation will consequentially bring on the occurrence of quantum effect. So many difficulties must be solved from some new angles.

On the other hand, as one of the most resplendent achievements in the 20[th] century, quantum theory is more rapidly developing. Especially, many results of quantum information technology have shown that quantum computer can effectively increase efficiency for solving some important classical problems and it even can solve some hard problems that classical computer can't solve; quantum communication can realize high-precision secrecy communication and increase capacity of channel [9]. So manipulation of quantum systems and applications of quantum effect become a most important research point for many scientists. Here, we consider the fusion of quantum theory and robot technology, and use quantum system to construct a new-type robot—quantum robot. The conception of quantum robot has been presented from physics by Benioff in 1998 [10,11], where he emphasized the importance of quantum computer in quantum robot and the quantum robot isn't aware of its environment, doesn't make decisions, doesn't carry out experiments or make measurement. However, we give a structure of quantum robot from engineering and consider the information exchange and learning control of quantum robot. Since quantum robot applies quantum effect, it solves the difficulties resulting from micromation. Moreover, the performance of sensors can be improved through equipping quantum robot with quantum sensors [12-16], and the speed of robot learning and behavior decision can be increased using powerful parallel computing, fast searching ability and efficient learning of quantum algorithms.

The organization of this paper is as follows. Section 2 presents a system structure of quantum robot and describes the functions of three fundamental parts including MQCU, quantum controller/actuator and information acquisition units. In Section 3 we use Grover algorithm to searching problem of quantum robot and also propose a novel machine learning algorithm—quantum reinforcement learning (QRL) for quantum robot. The theoretic results show that quantum robot can reduce the complexity of O($N^2$) in traditional robot to O($N\sqrt{N}$) using Grover algorithm, and the simulation results demonstrate that quantum robot is also superior to traditional robot in efficient learning by QRL algorithm. Section 4 compares quantum robot with classical robot and suggest some possible applications according to the advantages of quantum robot. Conclusion and remarks are given in Section 5.

## 2   System Structure of Quantum Robot

In 1998, Benioff firstly gave a conception of quantum robot where a quantum robot is described as a mobile quantum system that includes an on-board quantum computer and needed ancillary systems [10]. There, he emphasized the importance of quantum computer in quantum robot, and the robot described there has no awareness of its environment and doesn't make decisions or measurements. Here, we give an alternative definition considering quantum robot sensing and processing information of its external environment. *Quantum robot is a mobile physical apparatus designed using quantum effect of quantum system, which can sense the environment and its own state, and can also process quantum information and accomplish some tasks*. A quantum robot system includes three interactional parts (Fig.1.): MQCU (multi quantum computing units), quantum controller and actuator, and information acquisition units. A detailed description of each part is given as follows.

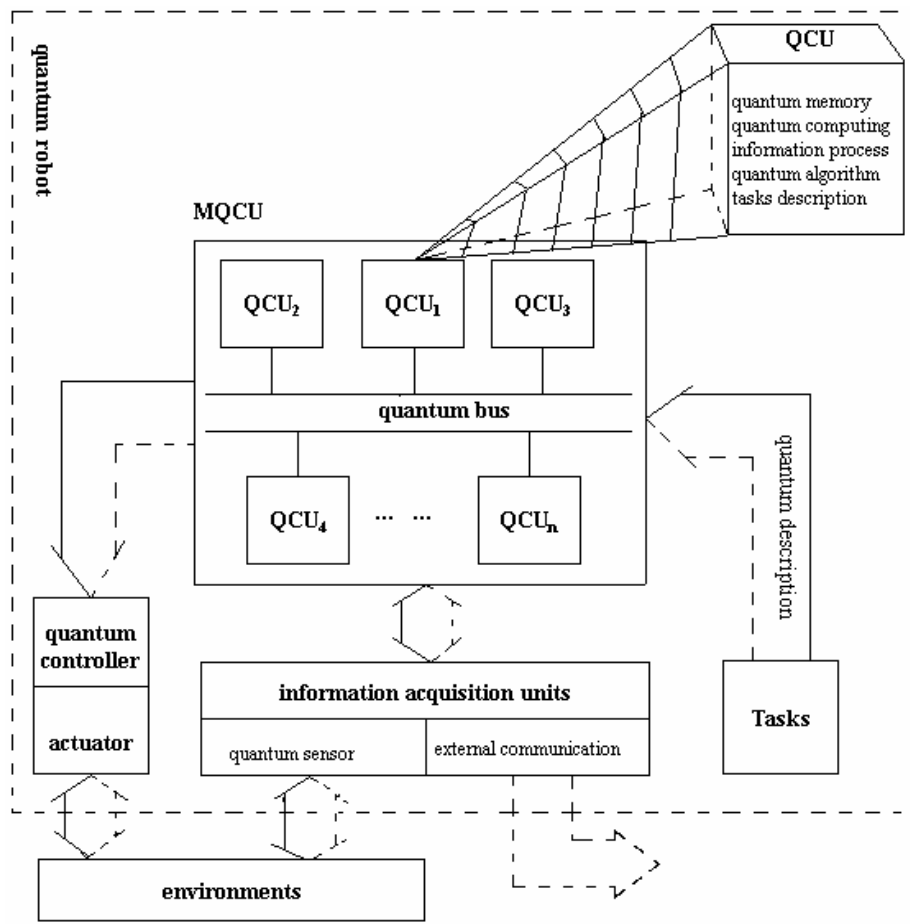

**Fig. 1.** System structure of quantum robot. A quantum robot includes three fundamental parts: MQCU, quantum controller and actuator, and information acquisition units.

**MQCU.** MQCU is the information processing center and acts as the cerebrum of quantum robot. It receives some tasks described using quantum language and exchanges information with its environment via quantum sensors or external communication. By storing, analyzing, computing and processing all kinds of information including task information, environment information and sensing information, the cerebrum can construct some appropriate quantum control algorithms and export indication signal into quantum controller to tell actuator what operation should be carried out. Usually, MQCU is made up of many QCUs (quantum computing units), each QCU accomplishes some specific tasks and exchanges information with each other through quantum bus. And quantum bus may be a refreshable entanglement resource [17] or some other quantum circuits. Besides some general functions of classical bus such as CAN bus, PCI bus and PC104 bus, quantum bus should also be able to exchange or preprocess some quantum information. According to quantum information theory, the QCUs can exchange information with each other more rapidly and secretly than MCUs of classical robot. Here, each QCU is a quantum information processor and it can perform some concrete task such as quantum computing, quantum memory and task description. In actual application, quantum computer can be used as the main part of QCU.

Quantum computer is a physical apparatus which can process quantum information and perform parallel computation by manipulating quantum state in a controlled way. In quantum computer, information unit (called quantum bit, or qubit) can lie in the coherent superposition state of logical state 0 and 1, that is to say, it can simultaneously store 0 and 1, so it can effectively speed up the solutions of some classical problems, and even solve some hard problems that classical computer can't solve. The essential characteristics of quantum computation are quantum superposition and quantum coherence. With the rapid development of quantum computation technology, some quantum computer models can be constructed using NMR, ion traps and photons. In the present robot system, quantum computer can act as QCU and accomplish some specific tasks such as storing, analyzing, computing and processing all kinds of information to help MQCU construct some appropriate quantum control algorithms.

**Quantum controller and actuator.** They are the execution and control apparatus of quantum robot. Quantum controller receives and processes indication signal from MQCU and informs actuator to carry out corresponding actions. It acts as the connection between the cerebrum (MQCU) and arm (actuator) of quantum robot. Commonly, quantum controller is a quantum system, such as quantum CNOT gate. Moreover, one may design useful quantum controller under the direction of rapidly developing quantum control theory [18-23].

The actuator executes some actions determined by the indication signal from quantum controller. An actuator may be a pure quantum system or a semiclassical apparatus capable of processing quantum information. Generally, the actuator can exchange quantum information as well as classical information with quantum controller. In some specific circumstances, the actuator and quantum controller can be looked upon as an apparatus, which can directly receive information from MQCU and carry out some actions on its environments. Besides quantum sensor, the actuator is another interaction channel between quantum robot and its environment.

**Information acquisition units.** As the same as traditional robot, quantum robot need also perceive its environments and acquire information through information acquisition units. In the present robot, quantum sensor perceives the information of its environments and the robot may also receive some other information from distant mainframe or other quantum robots through external communication unit. Usually, the information to be acquired includes quantum information and classical information. However, according to quantum theory, the acquisition of quantum information is difficult since quantum measurement destroys the quantum state of system. So quantum nondemolition (QND) measurement is an important task in quantum robot. In the robot system, quantum sensor acts as a key role for information acquisition. Quantum sensor is a kind of microstructure sensor, which is designed by applying quantum effect. To classical faint signal, nowadays two kinds of quantum sensors, superconduction quantum interference device (SQUID) sensor [13,14] and quantum well Hall sensor [15, 16], can be used in the quantum robot.

SQUID sensor is extremely sensitive magnetic sensor that is based on the principles of superconductivity, the Meissner effect, flux quantisation and the Josephson effect. Using Josephson effect, SQUID sensor can convert minute changes in current or magnetic field to a measurable voltage and detect magnetic fields as small as $10^{-10}$ Tesla. Quantum well Hall sensor is a kind of high-performance micro Hall sensor and uses two-dimensional electron gases to obtain a compromise between high mobility and high carrier concentration while maintaining a reasonably high sheet resistance. For example, we can construct a quantum well Hall sensor through sandwiching thin InAs layers between AlGaSb layers and the sensor has high magnetic sensitivity and very excellent temperature stability as a result of a good confinement of two dimensional electron gases in quantum well structure. So quantum well Hall sensor can be used to detect faint electric magnetic field under different kinds of circumstances. Using high sensitivity and good temperature stability of quantum sensors, we can equip the present quantum robot with them to measure extremely weak electromagnetic field. Simultaneously, scientists are studying other new-types of quantum sensors which can acquire quantum information. Once they are realized, we may also equip quantum robot with them to sense all kinds of quantum signals and feed back them into MQCU.

Moreover, the quantum robot has a communication interface to exchange information with distant mainframe or other quantum robots, which can constitute a multi quantum robot system (Fig.2.). In external communication, quantum information can be exchanged and the advantages of quantum communication such as high information channel capability, perfect security and quantum teleportation can be fully used.

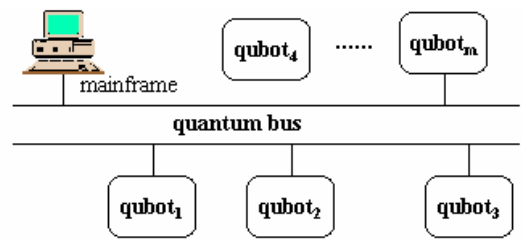

**Fig. 2.** Multi quantum robot system (where qubot, i.e. quantum robot).

Based on the above structure, it is obvious that quantum robot is also a kind of robot system which has the ability to accomplish certain tasks through perceiving environments with sensors and acting upon those environments with actuators. The characteristics rest with the physical implementation and their particular process of information. Suppose the task of quantum robot is to assist doctor with medical service in biomedicine, the information of task decomposition described by quantum language is sent to MQCU, moreover MQCU also receives some sensing information from external environments, and every QCU is in charge of some specific work such as track, navigation, estimation, diagnosis, etc. According to the processing results of MQCU, quantum controller obtains indi-cation signal from MQCU and informs actuator to carry out corresponding actions on external environments. Repeatedly, information acquisition units perceive environments and send sensing information into MQCU, MQCU processes the information and exports new signal or learning control algorithm to quantum controller and actuator until the task is accomplished. In the process, the design of learning control algorithm is a key aspect. Considering the characteristics of quantum robot, we will present two algorithms: quantum Grover algorithm [24] for robot searching and quantum reinforcement learning algorithm for learning of quantum robot.

## 3 Learning Control Algorithms for Quantum Robot

Just like traditional robot, a key aspect of quantum robot is also to design high-efficiency learning control algorithm. To traditional robot, it is difficult to perform some algorithms with high complexity and satisfy the requirement for real-time process because CPU cannot compute fast enough and the integration of chip is limited. However, quantum robot is essentially made up of many quantum systems and MQCU lies in the central position of whole quantum robot system, so one can use the ability of quantum parallel processing to design corresponding learning control algorithm, which can effectively reduce the complexity of solving problem and speed up information processing. So after the introduction of fundamental concepts of quantum parallel computation in subsection 3.1, we describe the Grover algorithms for searching problems of mobile robot and propose a novel quantum reinforcement learning algorithm.

### 3.1 Parallel Processing of Quantum Robot

Quantum robot is essentially a complex quantum system and its state is also represented with quantum state. To conveniently process all kinds of information, here we consider encoding all information according to qubits. In quantum information theory, the state of arbitrary qubit can be written into a superposition state as follows:

$$|\psi\rangle = \alpha |0\rangle + \beta |1\rangle \qquad (1)$$

where $\alpha$ and $\beta$ are complex coefficients and satisfy $|\alpha|^2 + |\beta|^2 = 1$. $|0\rangle$ and $|1\rangle$ are two orthogonal states (also called eigenstates of quantum state $|\psi\rangle$), and they correspond to logic states 0

and 1. $|\alpha|^2$ represents the occurrence probability of $|0\rangle$ when this qubit is measured, and $|\beta|^2$ is the probability of obtaining result $|1\rangle$. The value of classical bit is either Boolean value 0 or 1, but qubit can be prepared in the coherent superposition state of 0 and 1, i.e. qubit can simultaneously store 0 and 1, which is one of the main differences between traditional robot and quantum robot.

According to quantum information theory, the quantum computing process can be looked upon as a unitary transformation $U$ from input qubit to output qubit and $U$ is also a linear transformation. Since quantum robot is essentially a quantum system and MQCU lies in the central position of quantum robot, some process of quantum robot can also be looked upon as corresponding transformation. If one applies a transformation $U$ to a superposition state, the transformation will act on all eigenstates of this superposition state and the output will be a new superposition state by superposing the results of eigenstates. So when quantum robot processes function $f(x)$ by the method, the transformation $U$ can simultaneously work out many different results for a certain input $x$. This is analogous with classical parallel processing, so we call it parallel processing of quantum robot. And the ability of strong parallel processing is an important advantage of quantum robot over traditional robot.

Consider an $n$-qubit cluster which lies in the following superposition state:

$$|\psi\rangle = \sum_{x=00\cdots0}^{11\cdots1} C_x |x\rangle, \quad \sum_{x=00\cdots0}^{11\cdots1} |C_x|^2 = 1 \tag{2}$$

where the length of $x$ is $n$, $C_x$ is complex coefficients and $|C_x|^2$ represents occurrence probability of $|x\rangle$ measuring state $|\psi\rangle$. $|x\rangle$ has $2^n$ values, so the superposition state can be looked upon as the superposition state of all integers from 1 to $2^n$. Since $U$ is linear, processing function $f(x)$ can give out:

$$U \sum_{x=00\cdots0}^{11\cdots1} C_x |x,0\rangle = \sum_{x=00\cdots0}^{11\cdots1} C_x U |x,0\rangle = \sum_{x=00\cdots0}^{11\cdots1} C_x |x, f(x)\rangle \tag{3}$$

where $|x,0\rangle$ represents the input joint state and $|x, f(x)\rangle$ is the output joint state. Based on the above analysis, it is easy to find that an $n$-qubit cluster can simultaneously process $2^n$ states. However, this is different from classical parallel processing as quantum parallel processing doesn't necessarily make a tradeoff between processing time and needed physical space. In fact, it provides an exponen-tial-scale processing space in the *n*-qubit linear physical space. So quantum robot can effectively speed up information process and faster solve some problems such as navigation and decision.

### 3.2 Searching Algorithm of Quantum Robot

Just as traditional robot, most planning and control problems of quantum robot can also come down to searching problems. Thus we will put forth an abstract robot planning problem, and apply quantum Grover algorithm to it as an example.

First let's consider a robot planning system whose state evolves according to certain transition probabilities that depend on a control $u$. If the robot is in state $i$ and chooses control $u$ according to a policy $\pi$, it will move to state $j$ with probability $p_{ij}(u)$ and a cost $g(i,u,j)$. Now suppose that the desirability of a state is defined as $V$, which means the value of a state. The task of robot planning system is to find out the optimal sequence of $V(state)$, which satisfies some forms of Bellman's equation

$$V^*(i) = \min_u E[g(i,u,j) + V^*(j) | i, u] \quad \text{(for all } i, j\text{)} \quad (4)$$

where $E[\cdot | i, u]$ is the expected value. So at a certain state $i$, the robot planning problem is simply an unstructured searching problem to find the action $a_i$ (or the next state $j$) which is the optimal.

As for unstructured searching problems of searching space *N*, algorithm complexity is O($N$) in classical computation. In robotics, it is an important task to search suitable action from action space based on the current state of robot. If the complexity of state (or action) space is O($N$), the problem complexity in traditional robot is O($N^2$), but quantum robot can reduce the complexity to O($N\sqrt{N}$) by using Grover searching algorithm.

If there are *N* actions ($2^n \geq N \geq 2^{n-1}$), express them with *n* qubits:

$$|\psi_0\rangle = \sum_{l=00\cdots0}^{11\cdots1} a_l |l\rangle \quad (5)$$

where the length of *l* is *n*. For convenience, formula (5) can also be expressed as

$$|\psi_0\rangle = \sum_{i=1}^{2^n} a_i |i\rangle \quad (6)$$

Assume the cell to be found corresponds to $|k\rangle$, now we use quantum Grover algorithm to complete the searching task.

At first, we prepare a quantum state

$$|s\rangle = \frac{1}{\sqrt{2^n}} \sum_{i=1}^{2^n} |i\rangle \tag{7}$$

which is an equal weight superposition state. This can be accomplished by applying the Hadamard transformation to each qubit of the $n$-qubit state $|00\cdots00\rangle$ [9, 25]. Then we construct a reflection transform

$$U_s = 2|s\rangle\langle s| - I \tag{8}$$

Geometrically, when $U_s$ acts on an arbitrary vector, it preserves the component along $|s\rangle$ and flips the component in the hyperplane orthogonal to $|s\rangle$. Thus if apply $U_s$ to $|\psi_0\rangle$, we get

$$U_s|\psi_0\rangle = 2|s\rangle\langle s|\psi_0\rangle - |\psi_0\rangle = 2|s\rangle\sqrt{2^n}\langle a\rangle - |\psi_0\rangle$$

$$= 2\frac{1}{\sqrt{2^n}}\sum_{i=1}^{2^n}|i\rangle\sqrt{2^n}\langle a\rangle - \sum_{i=1}^{2^n}a_i|i\rangle = \sum_{i=1}^{2^n}(2\langle a\rangle - a_i)|i\rangle \tag{9}$$

where

$$\langle a\rangle = \frac{1}{2^n}\sum_{i=1}^{2^n}a_i \tag{10}$$

Now we give another reflection transform

$$U_k = -2|k\rangle\langle k| + I \tag{11}$$

where $|k\rangle$ is the $k$-th eigenstate and by applying it to state $|\psi_0\rangle$, we obtain

$$U_k|\psi_0\rangle = -2|k\rangle\langle k|\psi_0\rangle + |\psi_0\rangle = -2|k\rangle a_k + |\psi_0\rangle = \sum_{i=1, i\neq k}^{2^n} a_i|i\rangle - a_k|k\rangle \tag{12}$$

It is easy to see that $U_k$ only changes the amplitude's sign of $|k\rangle$ in the superposition state. Thus we can form a unitary transformation (Grover iteration) [9, 24]

$$U_G = U_s U_k \tag{13}$$

By repeatedly applying the transformation $U_G$ on $|\psi_0\rangle$, we can enhance the probability amplitude of $|k\rangle$ while suppressing the amplitude of all the other states $|i \neq k\rangle$ [24]. Iterating the transformation enough times, we can perform a measurement on the system to make the state $|\psi_0\rangle$ collapse into $|k\rangle$ with a probability of almost 1.

Define angle $\theta$ satisfying $\sin^2\theta = 1/2^n$. After applying the $U_G$ $j$ times to $|\psi_0\rangle$, the amplitude of $|k\rangle$ will become

$$a_k^j = \sin((2j+1)\theta) \tag{14}$$

If $j = (\pi - 2\theta)/4\theta$, then $(2j+1)\theta = \pi/2$ and $a_k^j = 1$. However, we must perform an integer number of iterations. Boyer has shown in [26] that the probability of failure is no more than $1/2^N$ if we perform the Grover iteration $\mathrm{int}(\pi/4\theta)$ times (here the function $\mathrm{int}(x)$ return the integer part of $x$). According to Grover algorithm, when $N$ is large, quantum robot can find the action corresponding to $|k\rangle$ with a high probability of [1-O($1/N$)]. Since Grover algorithm can find needed result with almost 1 probability in $\sqrt{N}$ steps, quantum robot reduces the complexity and can find suitable action from action space based on the current state of robot. If the number of states is $10^4$, considering different number of actions, the problem complexities in traditional robot and quantum robot can be compared as Table 1. From the Table, we can find that the advantage of quantum robot is more and more prominent with the increase of number of actions.

**Table 1.** Comparisons of problem complexities in traditional robot and quantum robot

| Number of actions | $10^2$ | $10^3$ | $10^4$ | $10^5$ | $10^6$ | $10^7$ | $10^8$ |
|---|---|---|---|---|---|---|---|
| Complexity in traditional robot | $10^6$ | $10^7$ | $10^8$ | $10^9$ | $10^{10}$ | $10^{11}$ | $10^{12}$ |
| Complexity in quantum robot | $10^5$ | $3.2\times 10^5$ | $10^6$ | $3.2\times 10^6$ | $10^7$ | $3.2\times 10^7$ | $10^8$ |

### 3.3 Learning Algorithm of Quantum Robot

The essence of robot learning is to deal with state-action pair {State(t), Action(t)} [31]. Learning methods are generally classified into supervised, unsupervised and reinforcement learning (RL). Supervised learning requires explicit feedback provided by input-output pairs and gives a map from input to output. And unsupervised learning only processes on the input data. However, RL uses a scalar value named reward to evaluate the input-output pairs and learns by interaction with environment through trial-and-error. Since 1980s, RL has become an important approach to machine intelligence [27-30], and is widely used in artificial intelligence, especially in robot [32- 34], due to its good performance of on-line adaptation and powerful learning ability of complex nonlinear system [27, 28, 35]. To adapt learning algorithm to quantum robot, we propose a novel learning method--quantum reinforcement learning (QRL) inspired by the quantum superposition and quantum parallelism.

Let $N_s$ and $N_a$ be the number of states and actions, then choose numbers $m$ and $n$, which are characterized by the following inequations:

$$N_s \leq 2^m \leq 2N_s, N_a \leq 2^n \leq 2N_a \qquad (15)$$

Use $m$ and $n$ qubits to represent state set $S = \{s\}$ and action set $A = \{a\}$ respectively:

$$s: \begin{bmatrix} a_1 & a_2 & \cdots & a_m \\ b_1 & b_2 & & b_m \end{bmatrix}, \text{where } |a_i|^2 + |b_i|^2 = 1, \ i = 1,2,...m$$

$$a: \begin{bmatrix} \alpha_1 & \alpha_2 & \cdots & \alpha_n \\ \beta_1 & \beta_2 & & \beta_n \end{bmatrix}, \text{where } |\alpha_i|^2 + |\beta_i|^2 = 1, \ i = 1,2,...n$$

Thus they may be in superposition state:

$$|s^{(m)}\rangle = \sum_{s=00\cdots0}^{11\cdots1} C_s |s\rangle, \ |a^{(n)}\rangle = \sum_{a=00\cdots0}^{11\cdots1} C_a |a\rangle \qquad (16)$$

where $C_s = x_s + iy_s$ and $C_a = u_a + iv_a$ are complex numbers. The mapping from states to actions is $f(s) = \pi : S \rightarrow A$ and we will get:

$$f(s) = |a_s^n\rangle = \sum_{a=00\cdots0}^{11\cdots1} C_a |a\rangle \qquad (17)$$

$|C_a|^2$ denotes the probability of $|a\rangle$ when $|a_s^n\rangle$ is measured. Based on the above express, the procedural form of QRL can be described as follows.

**Procedure QRL**

Initialize $|s^{(m)}\rangle = \sum_{s=00\cdots0}^{11\cdots1} C_s |s\rangle$, $f(s) = |a_s^n\rangle = \sum_{a=00\cdots0}^{11\cdots1} C_a |a\rangle$ and $V(s)$ arbitrarily

Repeat (for each episode)

For all states $|s^{(m)}\rangle = \sum_{s=00\cdots0}^{11\cdots1} C_s |s\rangle$:

(1) Observe $f(s)$ and get $|a\rangle$;

(2) Take action $|a\rangle$, observe next state $|s'^{(m)}\rangle$, reward $r$

Then update: $V(s) \leftarrow V(s) + \alpha(r + \gamma V(s') - V(s))$

$$C_a \leftarrow e^{\lambda(r+V(s'))} C_a$$

Until for all states $|\Delta V(s)| \leq \varepsilon$.

QRL is inspired by the state superposition principle of quantum state and quantum parallel computation. The state value can be represented with quantum state and be obtained by randomly observing the simulated quantum state, which will lead to state collapse according to quantum measurement postulate. And the occurrence probability of eigenvalue is denoted by probability amplitude, which is updated according to rewards. So this approach represents the whole state-action space with the superposition of quantum state and makes a good tradeoff between exploration and exploitation using probability. What's more, the representation method is consistent with quantum parallel computation and can speed up learning dramatically.

In [27], Bertsekas and Tsitsiklis have verified that stochastic iterative algorithms, under certain exploration policy, converge at the optimal state value function $V(s)^*$ with probability 1 when the following hold (where $\alpha_k$ is stepsize):

$$\lim_{T\to\infty}\sum_{k=1}^{T}\alpha_k = \infty, \qquad \lim_{T\to\infty}\sum_{k=1}^{T}\alpha_k^2 < \infty \tag{18}$$

And QRL is the same as traditional RL, but differences lie in: (1) exploration policy is based on the collapse theory of quantum measurement while being observed; (2) parallel computation. So the modification of RL does not affect the characteristic of convergence and QRL algorithms converge when (18) holds.

To evaluate QRL algorithm in practice, consider the typical rooms with corridor example, gridworld environment of four rooms and surrounding corridors as shown in Fig. 3. Each cell of the grid corresponds to an individual state of the environment. From any state the robot (or agent) can perform one of four primary actions: up, down, left and right, and actions that would lead into a blocked cell are not executed. The task of the algorithms is to find an optimal policy which will let the robot move from *S* to *G* with minimized cost (or maximized rewards). The simulation environment is based on Windows 2000 and Visual C++, and the results are processed with Matlab 6.5.

**Experimental set-up.** In QRL, the action selecting policy is obviously different from traditional RL algorithms, which is inspired by the collapse theory of quantum measurement. And probability amplitudes $|C_a|^2$ is used to denote the probability of an action defined as $f(s) = |a_s^n\rangle = \sum_{a=00\cdots 0}^{11\cdots 1} C_a |a\rangle$. For the four cell-to-cell actions $|C_a|^2$ is initialized uniformly. In this $13 \times 13 (0 \sim 12)$ grid world, the initial state and the goal are cell(4,4) and cell(8,8) respectively.

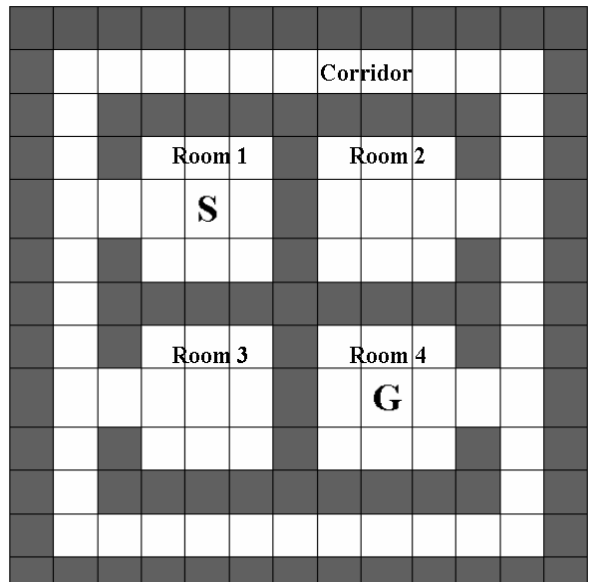

**Fig. 3.** The example of rooms with corridor is a gridworld environment with cell-to-cell actions (up, down, left and right). The labels *S* and *G* indicate the initial state and the goal in the simulation experiment described in the text.

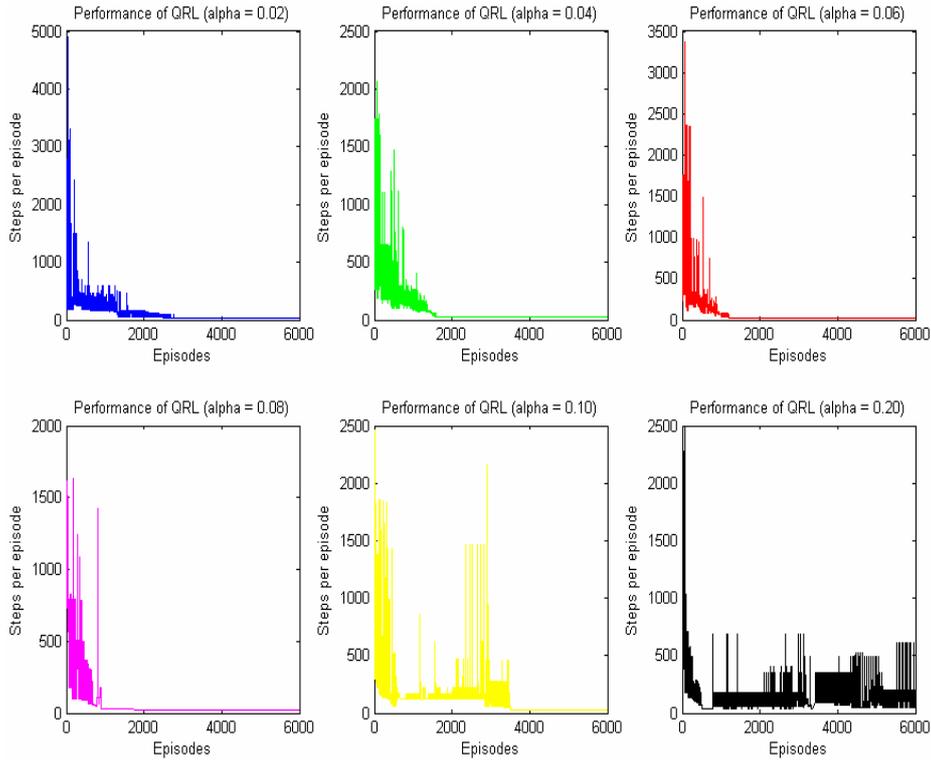

**Fig. 4.** Performance of QRL in the example of rooms with corridor.

**Results and analysis.** Learning performance for QRL is plotted in Fig. 4. We observe that given a proper stepsize (alpha < 0.10) this algorithm learns extraordinarily fast at the beginning phase, and then steadily converges to the optimal policy that costs 25 steps to the goal *G*. As the stepsize increases from 0.02 to 0.10, this algorithm learns faster but more unsteadily. When the stepsize is 0.20, it cannot converge to the optimal policy. The results show that QRL algorithm excels other RL algorithms in the following two main aspects: (1) Action selecting policy makes a good tradeoff between exploration and exploitation using probability, which speeds up the learning and guarantees the searching over the whole state-action space as well. (2) Updating is carried through parallel, which will be much more prominent in the future when practical quantum apparatus comes into use instead of being simulated on traditional computers.

## 4 Applications of Quantum Robot

Before presenting the potential applications of quantum robot, we firstly compare quantum robot with traditional robot. Since quantum robot applies quantum effect and can process quantum information, there are many differences between quantum robot and traditional robot (Table 2).

**Table 2.** Comparisons between quantum robot and traditional robot

| Compared items | Quantum robot | Traditional robot |
|---|---|---|
| System property | Quantum system | Mechanical system |
| On-board Sensors | Quantum sensors such as SQUID sensors and quantum well Hall sensors | Infrared sensors, ultrasonic sensors, CCD camera and etc. |
| Law to obey | Quantum mechanical law | Classical mechanical law |
| Information processing centre | MQCU (Multi quantum computing units) | Classical processor such as classical computer |
| Information | Quantum information and classical information | Classical information |
| Delicacy | High | Relatively lower |
| Space size | Microcosmic | Mostly macro |
| Technical difficulty | More difficult | Easier |
| Communication manner | Quantum communication and classical communication | Classical communication |
| Parallel process ability | Powerful | Weak |

Thus it can be seen that quantum robot as a brand-new robot can solve the difficulties resulting from micromation. It makes use of quantum sensor to acquire quantum information as well as classical information from its environments and has extremely high delicacy, which effectively overcomes the limitation of existing sensors' performance. At the same time, it can directly use the advantages of quantum information technology to speed up the control and decision of robot. So it will have wide application prospect in the fields of military affairs, national defense, aviation and spaceflight, biomedicine, science research, safety engineering and other daily life.

For example, it can be used as patrol warrior in military affairs and national defense. If we have measured in advance the magnetic fields near important ports and military bases, using the high delicacy of quantum sensor, quantum robot may perceive the faint change of magnetic field resulting from the closing of nuke and scout to effectively forewarn decision-maker. In aviation and spaceflight, it can be used to design Mars detector and moon vehicle to accomplish space exploration through using its high delicacy of perceiving environments, powerful ability of processing information and more secure communication. In biomedicine, it can be used to patient with a contagious disease for examining or tracking state of an illness, and acts as SARS nurse and contagion doctor. Once quantum robot is successfully constructed, considering the minisize characteristic, it will be possible to establish molecule-scale medicinal robot. Thus quantum robot can be injected into body, moves along with blood circulation, and detects potential pathological changes in body. At the same time, it will likely provide a brand-new way to study life. In science research, it is possible to use quantum robot to solve most complex problems with less physical resources and provide a test-bed for experiment research of physics, chemistry and information so that some experiment realization of quantum communication, quantum computing and quantum control will possibly become easier. In safety engineering, one may take advantage of the high delicacy and small size of quantum robot to many aspects such as anti-terror forewarning, fire forecasting, guarding against theft and traffic directing.

# 5 Conclusion

With the advancement of technology, robots have been widely applied to many fields. At the same time, some new-type robots such as nanorobot, biorobot and medicinal robot also come to our world. This paper proposes another new-type robot—quantum robot, by fusing quantum theory into robot technology. A quantum robot is a mobile physical apparatus designed using quantum effect of quantum system, which can sense the environment and its own state, and can also process quantum information and accomplish some tasks. It should be composed of three fundamental parts: MQCU, quantum controller and actuator, and information acquisition units. To adapt some algorithm to the system structure of quantum robot, quantum searching algorithm is presented and quantum reinforcement learning is proposed for quantum robot. The theoretic results show that quantum robot can reduce the complexity of $O(N^2)$ in traditional robot to $O(N\sqrt{N})$ using quantum Grover algorithm, and the simulation experiments demonstrate that quantum robot is also superior to traditional robot in efficient learning since quantum reinforcement learning makes a good tradeoff between exploration and exploitation using probability. So it has many potential important applications in military affairs, national defense, aviation and spaceflight, biomedicine, science research, safety engineering and other daily life.

To implement a real quantum robot is no doubt a very challenging mission, which consists of three kinds of work: (1) synthesis architecture for quantum robot systems; (2) physical implementation including computing units, sensors, actuators and communication hardware; (3) software level researches such as related theories and programming issues. Though our work is only the first step to practical quantum robot, it has luciferous future with the rapid development and gradual maturation of quantum information technology and quantum control theory.